\documentclass[letterpaper]{article} % DO NOT CHANGE THIS
\usepackage{aaai24}  % DO NOT CHANGE THIS
\usepackage{times}  % DO NOT CHANGE THIS
\usepackage{helvet}  % DO NOT CHANGE THIS
\usepackage{courier}  % DO NOT CHANGE THIS
\usepackage[hyphens]{url}  % DO NOT CHANGE THIS
\usepackage{graphicx} % DO NOT CHANGE THIS
\usepackage{natbib}  % DO NOT CHANGE THIS AND DO NOT ADD ANY OPTIONS TO IT
\usepackage{caption} % DO NOT CHANGE THIS AND DO NOT ADD ANY OPTIONS TO IT
\usepackage{algorithm}
\usepackage{algorithmic}

\usepackage{newfloat}
\usepackage{listings}
\DeclareCaptionStyle{ruled}{labelfont=normalfont,labelsep=colon,strut=off} % DO NOT CHANGE THIS
\floatstyle{ruled}
\newfloat{listing}{tb}{lst}{}
\floatname{listing}{Listing}
 \nocopyright% -- Your paper will not be published if you use this command
\newcommand{\primarykeywords}{
(1) Applications;
(2) Learning;
}

\usepackage[switch, modulo]{lineno}
\title{On Automating Video Game Regression Testing by Planning and Learning}
\author {
    Tom\'a\v{s} Balyo,
    G. Michael Youngblood,
    Filip Dvo\v{r}\'ak, 
    Luk\'a\v{s} Chrpa,
    and
    Roman Bart\'ak
}
\affiliations {
    Filuta AI, Inc., 199 Water St Fl 34, New York, NY 10038\\
    \{tomas, michael, filip, lukas, roman\}@filuta.ai
}

\begin{document}

\maketitle

\begin{abstract}

In this paper, we propose a method and workflow for automating regression testing of certain video game aspects using automated planning and incremental action model learning techniques. The basic idea is to use detailed game logs and incremental action model learning techniques to maintain a formal model in the planning domain description language (PDDL) of the gameplay mechanics.
The workflow enables efficient cooperation of game developers without any experience with PDDL or other formal systems and a person experienced with PDDL modeling but no game development skills.
We describe the method and workflow in general and then demonstrate it on a concrete proof-of-concept example --- a simple role-playing game provided as one of the tutorial projects in the popular game development engine Unity.
This paper presents the first step towards minimizing or even eliminating the need for a modeling expert in the workflow, thus making automated planning accessible to a broader audience.
% Remember to add the key takeaway from the Conclusion as a result

\end{abstract}

\section{Introduction}
Game testing is a vital yet complex component of video game development, focusing on identifying and resolving bugs, performance issues, and ensuring a high-quality user experience. This multifaceted process involves testing diverse game scenarios, enhancing user interface and playability, and maintaining stability and efficiency. It also includes compliance with standards and regulations, especially for games with online components. Modern game testing combines automated and manual methods, often engaging professional testers and the gaming community, to not only fix issues but also elevate overall player satisfaction and game quality. However, automation of testing methods still has a long way to go, especially for techniques that do not significantly disrupt developer workflow \cite{politowski2022towards}. This paper focuses on a contribution to those automated methods by leveraging planning.

\emph{Planning} is a less common skill game developers utilize but can be a powerful tool for game testing. One key aspect of that difficulty comes from the authorial burden of writing domain and problem descriptions in PDDL (Planning Domain Definition Language) ~\cite{aeronautiques1998pddl}. Good PDDL modeling is a rare skill that develops over time, so an efficient use of modelers by enabling them to be more effective is another key to adoption. 

Our work addresses these critical issues to help game developers utilize planners for generating test scripts driven by PDDL through automation support. We accomplish this by taking the game logs, produced in software development, providing easy-to-implement guidance to augment the logging for action model capture, and performing automated domain synthesis to generate PDDL models easily validated, verified, and adjusted by the game developer. These good PDDL examples also improve the developers' understanding and authorial abilities in PDDL in the flow. Goal templates are then used to formulate domain problems of interest for testing. We demonstrate our approach and process on a simple RPG game in the ubiquitous Unity game engine ~\cite{unity}.

\section{Motivation}
We will focus on regression testing during game development, i.e., running frequent tests to ensure that recent changes did not introduce unintended breaks and the game still performs as expected. This is a particularly repetitive and dull task for a human tester but crucial for efficient development. It is often the case, that longer the time difference between the introduction and discovery of the bug, the harder it is to fix.

\subsection{What can be tested with Planning}
The most obvious and important usage is to test whether the game can be completed, i.e., each intended objective can be achieved from a starting state of each scenario. Next, we can generate a large amount of random test scenarios. This approach is beneficial if we can execute and evaluate test scenarios automatically. Furthermore, we can search for dead-ends in the game, i.e., states that can be reached but are impossible to continue from or to win. In modern game design, such states are most undesirable and annoying for the players. If the PDDL model is detailed enough, one can detect such dead ends by running planners from random reachable states.
Last but not least, one can prove that undesired shortcuts are impossible through regular gameplay using optimal planners.

\subsection{Applicability}
The classical planning based testing approach is most suitable for games with many discreet causal interactions such as role playing games (RPG), point-and-click adventures, strategy games, visual novels or puzzle games. It is less suitable for fast-paced action oriented games such as shooters or driving games. For such games, reinforcement learning based approaches, that have been studied in previous work, seem to better suitable (see the Related Work Section of this paper). Overall, we observe that the reinforcement learning and the planning based approaches can complement each other rather well.

\section{Preliminaries}
The planning problem is to find a plan --- a sequence of grounded actions that transform the world from an initial state to a goal state, i.e., a state where all goal conditions are satisfied.
A planning problem instance consists of a domain definition and a task definition. The domain definition describes possible actions with their preconditions and effects. In STRIPS planning~\cite{strips}, both preconditions and effects are sets of predicates and negated predicates connected by a conjunction.
The task definition contains the descriptions of the initial state and the goal conditions.

Planning action model learning (AML) involves synthesizing a domain definition from logs with state sequences. Many algorithms have been published for this task~\cite{observer,fama,LAMP,arms, sam}, a recent overview and new implementation of multiple techniques is provided in the MacQ project~\cite{macq}. For the evaluation in this paper we use our own new action learning method~\cite{balyo2024learning}.

\begin{figure*}
    \centering
    \includegraphics[width=\linewidth]{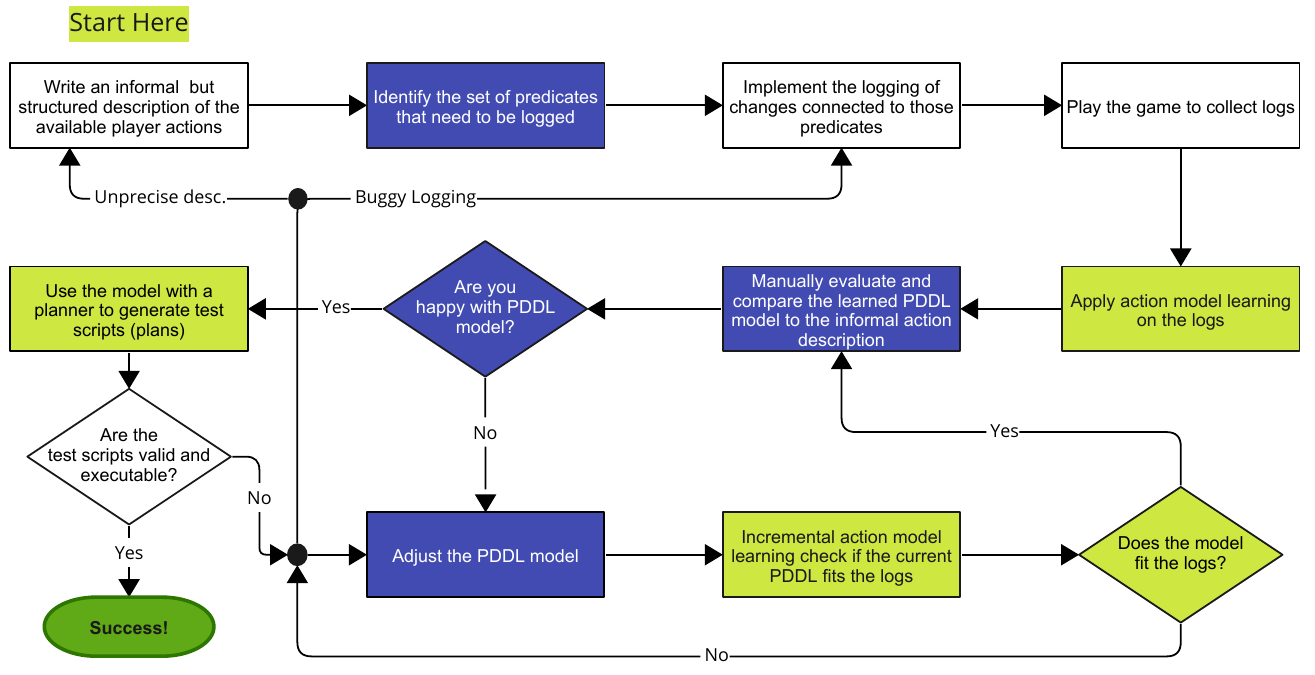}
    \caption{The overview of the workflow for acquiring PDDL domain models based on logs from the game execution.
    The objects with white background represent the tasks best done by game developer(s). The tasks and decisions with a blue background should be done by the PDDL modeller. The light green items can be done automatically.}
    \label{fig:worflow}
\end{figure*}

\begin{figure*}
\begin{lstlisting}[language=Lisp,frame=single]
...
2023-08-15 19:31:20 697 ;NEXT-STATE
2023-08-15 19:31:21 916 (not (questState apples-quest ready));start_quest
2023-08-15 19:31:21 916 (questState apples-quest started);start_quest
2023-08-15 19:31:21 917 ;NEXT-STATE
2023-08-15 19:31:24 315 (not (location player 8,-4));move
2023-08-15 19:31:24 315 (location player 9,-4);move
2023-08-15 19:31:24 316 ;NEXT-STATE
2023-08-15 19:31:24 574 (not (location player 9,-4));move
2023-08-15 19:31:24 574 (location player 10,-4);move
2023-08-15 19:31:24 574 ;NEXT-STATE
2023-08-15 19:31:24 598 (not (collected apples 0));pickup
2023-08-15 19:31:24 598 (collected apples 1);pickup
2023-08-15 19:31:24 598 (not (location golden-apple0 10,-4));pickup
2023-08-15 19:31:24 598 (location golden-apple0 none);pickup
2023-08-15 19:31:24 598 ;NEXT-STATE
...
2023-08-15 19:31:29 816 ;NEXT-STATE
2023-08-15 19:31:30 055 (not (questState apples-quest started));complete_quest
2023-08-15 19:31:30 056 (questState apples-quest complete);complete_quest
2023-08-15 19:31:30 056 ;NEXT-STATE
...
\end{lstlisting}
    \caption{A fragment of the log file created by playing the game and used for action model acquisition.}
    \label{fig:log}
\end{figure*}

\section{Related Work}

Planning is rooted in search and, as such, has had a long history with games. At the beginning of AI, the focus was on games like chess ~\cite{newell1972human} and checkers ~\cite{samuel1959some}, which initially relied on the search for solutions. Planning continued to dominate in the 80s and 90s with checkers and chess with Deep Blue ~\cite{hsu1990deep} and Chinook ~\cite{schaeffer1996chinook}. Agent architectures and production systems added value, and soon, planning started to add value in games like bridge ~\cite{smith1998computer} and the class of Real-Time Strategy (RTS) games ~\cite{chung2005MCPlan}. In ~\cite{duarte2020survey}, they survey the history of planning and learning in games, covering the spectrum as well as diving into the lineage of planning from search, minimax and alpha-beta pruning, hierarchical task networks, and Monte Carlo Tree search, through classical planning, rapidly-exploring random trees, case-based planning, and behavior trees. Most of the work is focused on creating AI-driven opponents~\cite{WurmanBKMS0CDE022}, which are sometimes used to play both sides for evaluation, AI training, and testing.

Automated testing with AI has been a rising research focus more recently with work that has focused on agent-based approaches that include navigation mesh pathfinding \cite{shirzadehhajimahmood2021using}, reinforcement learning agents for finding design and environmental defects ~\cite{ariyurek2019automated, ferdous2022towards}, reinforcement learning for load testing ~\cite{tufano2022using}, modeling of user interaction for boundary testing ~\cite{owen2016modelling}, search for test case generation ~\cite{10.1007/978-3-030-88106-1_5}, and search for automated play testing ~\cite{ferdous2021search}. Despite planning use in game AI, we do not see its use in game testing more broadly beyond back to search. However, as evidenced by Bram Ridder's (AI Programmer for Rebellion) keynote talk at the 2021 AIIDE Conference on ``Improved Automated Game Testing Using Domain-Independent AI Planning''~\cite{ImproveA39:online} and his 2021 GDC AI Summit talk ``Automated Game Testing Using a Numeric Domain Independent AI Planner,'' planning techniques for game testing are beginning to be used in the games industry mixed in with calls for more AI automation of testing~\cite{Automate82:online}.

\section{The General Workflow}

The general workflow is visualized in a flowchart in Figure~\ref{fig:worflow}. Each task requires a person (or a group) having one of the following two skill sets:
\begin{itemize}
    \item A game developer or a game designer (Developer for short). This is someone familiar with the rules of the game and the ability to modify the source code of the game to add log generation functionality. Also, they need to be able to play the game in such a way that all the available game mechanics are utilized. Lastly, they must be capable of (automatically) evaluating whether a given test script is consistent with the game's rules.
    \item A person with PDDL modeling skills (Modeller for short). This is someone who knows PDDL and has some experience with modeling planning scenarios in this language.
\end{itemize}

We start by writing an informal but structured description of all the legal player moves/actions. 
For each player action, we need to provide a name and list of preconditions (what conditions need to be valid in order to allow the player to play this action) and effects (what changes after the action is played out) written in natural language. This task should be easily doable by the developer. It requires no PDDL modeling experience, even though the required structure of the action is very similar to that of an STRIPS action.

Next, the informal action descriptions are handed to a modeler. The modeler's task at this point is only to identify a set of predicates required to formally describe all the properties mentioned in the preconditions and effects sections of the action descriptions.

Then, the set of predicates is given to a developer who needs to extend the game's source code to facilitate the logging of changes to those predicates. Now, we are ready to play the game while collecting logs. The logs are passed to an action model learning (AML) tool, and a PDDL model is automatically generated.

Many games already produce logs, so why cannot we just use existing logs instead of modifying the code and creating new ones? In principle, we could do that. However, in our experience, the existing logs (if they are available at all) usually only contain method calls and error messages and do not record changes of state variables. Those kinds of logs unfortunately cannot be used to learn action models because the required information (state changes) is not present in them.

Now, it is the modeler's turn again, as they need to examine and evaluate the generated PDDL model manually. Due to the imprecise nature of AML algorithms, the generated model often needs minor corrections and adjustments. After such adjustments, it is possible to verify automatically whether the new PDDL model is still consistent with the logs.
Note that the first model coming from the AML is always consistent with the logs (this is a property of the used action model learning algorithm~\cite{balyo2024learning}). If the model is consistent with the logs, the modeler can either continue editing or decide that it is good enough and can be passed to the planning step.
On the other hand, if the model is inconsistent with the logs, we need to either change the PDDL model or the logs. Changing the logs may be required because they are inconsistent due to bugs in the implementation of the logging process or because we are not logging the proper predicates. In the latter case, we may need to go back all the way to the beginning of the process and adjust the informal description, update the set of required predicates, and repeat all the follow-up steps (see Figure~\ref{fig:worflow}).

If the logs are consistent with the current domain model and the modeler believes it is accurate, we can use a planner to create a collection of plans. These plans are then handed over to the developers, who can check whether they are consistent with the game's rules following some simple goal templates. This process can be done manually or automatically if the game can execute test scripts. If no problems are discovered with the generated plans/test scripts, we are finished, and we have obtained a PDDL model representing the game's rules. If we discover any problems, we handle the situation the same way as in the case of having a domain inconsistent with the logs (see Figure~\ref{fig:worflow}).

\section{Proof of concept: A Simple RPG Game}

We will demonstrate our workflow on the tutorial project of ``Creator Kit: RPG''~\cite{rpg-demo}, available
in the popular video game engine Unity~\cite{unity} versions 2019.1 -- 2021.3.

It is a basic single-player RPG with a 2D environment viewed from above. The player controls a hero character
using the arrow keys. The hero can talk to non-player characters (NPCs) on the map. Some NPCs can provide a ``quest'' (a task for the hero to complete) when spoken to. The quests all follow the pattern of: ``collect $n$ items of type $T$''.
After a quest is activated, the relevant items appear on the map. To pick up an item, the hero only needs to walk near it. When the required amount of items is collected, the player must visit the NPC who provided the quest to complete it.
There are two quests in the demo: ``collect 3 golden apples'' and ``collect 10 chickens''.

Following the workflow defined above (in Figure~\ref{fig:worflow}), we need to start by providing an informal but
structured description of the possible actions. For our simple example, we only require four actions described in Table~\ref{tab:description}.

\begin{table}
    \centering
    \begin{tabular}{l l}
       \hline
       \multicolumn{2}{c}{Move hero}\\
       \hline
       Hero is at the start tile  &  Hero is at the goal tile \\
       Start and goal tiles \\
       are neighbours  & \\
       \hline
       \multicolumn{2}{c}{Pick up quest item}\\
       \hline
       Hero is at the same  &  The item is gone \\
       location as the item  & \\
       The quest related to & The hero has one more of \\
       this item is active & this kind of item  \\
       \hline
       \multicolumn{2}{c}{Start quest}\\
       \hline
       The quest is ready  & The quest is active \\
       Hero is at the same location & \\
       as the quest giver  &  \\
       \hline
       \multicolumn{2}{c}{Complete quest}\\
       \hline
       The quest is active  &  The quest is done \\
       The hero has enough items  & \\
       of the required type & \\
       Hero is at the same location &  \\
       as the quest giver &\\
       \hline
    \end{tabular}
    \caption{Structured informal description of player actions. The left column is the preconditions, right is the effects.}
    \label{tab:description}
\end{table}

\begin{figure*}
\begin{lstlisting}[language=Lisp,frame=single]
(define (domain rpg)
    (:requirements :strips :typing :negative-preconditions)
    (:types
        number state location locatable typename - object
        quest item hero - locatable
    )
    (:constants
    	active ready done - state
    )    
    (:predicates
        (next ?n1 - number ?n2 - number)
        (connected ?l1 - location ?l2 - location)
        (questState ?q - quest ?s - state)
        (itemType ?i - item ?t - typename)
        (questItemType ?q - quest ?t - typename)
        (location ?l - locatable ?loc - location)
        (collected ?t - typename ?n - number)
        (questItemCount ?q - quest ?n - number)
    )
    (:action start_quest
        :parameters(?quest - quest ?hero - hero ?location - location)
        :precondition (and
            (questState ?quest ready)
            (location ?hero ?location)
            (location ?quest ?location)
        )
        :effect (and
            (not (questState ?quest ready))
            (questState ?quest active)
        )
    )
    (:action complete_quest
        :parameters(?quest - quest ?hero - hero ?typename - typename 
                    ?number - number ?location - location)
        :precondition (and
            (questState ?quest active)
            (location ?hero ?location)
            (location ?quest ?location)
            (collected ?typename ?number)
            (questItemType ?quest ?typename)
            (questItemCount ?quest ?number)
        )
        :effect (and
            (not (questState ?quest active))
            (questState ?quest done)
        )
    )
\end{lstlisting}
    \caption{First half of the RPG domain PDDL description.}
    \label{fig:pddl1}
\end{figure*}

\begin{figure*}
\begin{lstlisting}[language=Lisp,frame=single]
    (:action pickup
        :parameters(?hero - hero ?item - item ?location - location 
                    ?number-1 ?number-2 - number ?typename - typename ?quest - quest)
        :precondition (and
            (collected ?typename ?number-1)
            (location ?item ?location)
            (location ?hero ?location)
            (itemType ?item ?typename)
            (questItemType ?quest ?typename)
            (questState ?quest active)
            (next ?number-1 ?number-2)
        )
        :effect (and
            (not (collected ?typename ?number-1))
            (not (location ?item ?location))
            (collected ?typename ?number-2)
        )
    )
    (:action move
        :parameters(?hero - hero ?location-1 ?location-2 - location)
        :precondition (and
            (location ?hero ?location-1)
            (connected ?location-1 ?location-2)
        )
        :effect (and
            (not (location ?hero ?location-1))
            (location ?hero ?location-2)
        )
    )
)
\end{lstlisting}
    \caption{Second half of the RPG domain PDDL description.}
    \label{fig:pddl2}
\end{figure*}

\begin{figure*}
\begin{lstlisting}[language=Lisp,frame=single]
(define (problem level1)
    (:domain rpg)
    (:objects
        player - hero
        ready - state
        apples chicken - typename
        apples-quest chicken-quest - quest
        chicken0 ... chicken9 - item
        golden-apple0 ... golden-apple2 - item
        n-1x-7 ... n9x2 - location
        n0 ... n9 - number
    )
    (:init
        (collected apples n0)
        (collected chicken n0)
        (questItemCount apples-quest n3)
        (questItemCount chicken-quest n10)
        (questItemType apples-quest apples)
        (questItemType chicken-quest chicken)
        (questState apples-quest ready)
        (questState chicken-quest ready)
        (location player n4x10)
        (location apples-quest n8x-4)
        (location chicken-quest n-6x7)
        (connected n-1x-7 n-2x-7)
        ...
        (itemType chicken0 chicken)
        (location chicken0 n-1x3)
        ...
        (itemType golden-apple0 apples)
        (location golden-apple0 n10x-4)
        ...
        (next n0 n1)
        ...
    )
    (:goal (and
        (questState apples-quest done)
        (questState chicken-quest done)
    ))
)
\end{lstlisting}
    \caption{The problem file for the RPG demo. Some of the objects and initial state predicates are redacted to shorten the listing.}
    \label{fig:pddl3}
\end{figure*}

\begin{figure*}
\begin{lstlisting}[language=Lisp,frame=single]
...
22: MOVE PLAYER N10X-4 N9X-4
23: MOVE PLAYER N9X-4 N8X-4
24: START_QUEST APPLES-QUEST PLAYER N8X-4
25: MOVE PLAYER N8X-4 N9X-4
26: MOVE PLAYER N9X-4 N10X-4
27: PICKUP PLAYER GOLDEN-APPLE0 N10X-4 N0 N1 APPLES APPLES-QUEST
28: MOVE PLAYER N10X-4 N10X-3
29: MOVE PLAYER N10X-3 N10X-2
30: MOVE PLAYER N10X-2 N10X-1
31: MOVE PLAYER N10X-1 N10X0
32: PICKUP PLAYER GOLDEN-APPLE1 N10X0 N1 N2 APPLES APPLES-QUEST
33: MOVE PLAYER N10X0 N11X0
34: MOVE PLAYER N11X0 N11X-1
35: PICKUP PLAYER GOLDEN-APPLE2 N11X-1 N2 N3 APPLES APPLES-QUEST
36: MOVE PLAYER N11X-1 N11X-2
37: MOVE PLAYER N11X-2 N10X-2
...
\end{lstlisting}
    \caption{A fragment of the plan to solve the demo level of the game.}
    \label{fig:plan}
\end{figure*}

In this example the modeller decided to use a grid-based (i.e., via tiles)  abstraction of locations\footnote{Using the grid abstraction turned out to be rather inconvenient. The reasons are
that it is not very robust if the items are not aligned to the grid and its is hard to determine which tiles are connected. In a game with free movement such as our example it would
be better to use a navigation-mesh-based abstraction~\cite{shirzadehhajimahmood2021using}.}
and came up with the following list of properties that should be logged:
\begin{itemize}
    \item location of the hero (tile),
    \item locations of each item (tile or none),
    \item the state of each quest (ready, active, done),
    \item the location of each quest-giving NPC (tile),
    \item number of items of each kind the hero has.
\end{itemize}
Additionally, we will need to know the following static properties of the environment and quests:
\begin{itemize}
    \item initial locations of each item (tile),
    \item type of each item (apple, chicken, ...),
    \item type of item required for a quest (apple, chicken, ...),
    \item number of items required to fulfill the quest,
    \item which tiles are neighbours.
\end{itemize}
To see the concrete predicate declarations corresponding to these properties see Figure~\ref{fig:pddl1}.

The developers then implemented logging and provided logs for learning (see Figure~\ref{fig:log}). The modeller refined the PDDL domain synthesized from the logs (see Figures~\ref{fig:pddl1}~and~\ref{fig:pddl2}) and used a planner to a generate a plan to win the game (see Figure~\ref{fig:plan}). We used the FF planner~\cite{hoffmann2001ff} which worked very well for our problem and found a plan in 0.01 seconds.

In order to test whether a generated plan (test script) is valid\footnote{A test script is valid if it is executable in a bug-free implementation of the game. If a test script is valid, but not executable, then a bug in the game is indicated}, we implemented automatic test script execution. 
In this case, it was very easy, since the game is controlled just by moving the hero.
Quests are activated/completed automatically when their requirements are fulfilled and the hero comes near their NPC.
Also, items are picked up automatically just by walking up to them. Therefore, to execute the test scripts
we only need to execute the Move actions. We did that by simulating the pushing of the arrow buttons in the directions of the next goal tile
until we are close enough to the center of that tile\footnote{A recording of the plan execution within the game is available here: \url{https://www.youtube.com/watch?v=BASKvQAbG04}}.

\section{Usage During Development}
The process we described should be applied in the beginning of the development process, when the game is simple and only contains a few game mechanics like our RPG demo. As the game is developed and new features are added, the PDDL model should be developed together with the game. Using incremental action model learning can aid the developers with maintaining the PDDL model.

\section{Challenges with Fully Automated Logging for Learning}

Deciding which properties/predicates should be logged and then implementing the logging into the game code can be complex and time-consuming. Therefore, automating this part of the workflow would be very beneficial. This section will discuss some examples of issues and challenges that need to be solved to achieve full automation.

In general, logging must be detailed enough to capture all the game mechanics precisely; therefore, it is necessary to log the values and changes of all variables, arrays, and other data structures of selected game-play related classes in the code. On the other hand, for automated planning to be efficient, we must work with an abstraction of the game rules. We need high-level (of abstraction) predicates describing world states for this. The level of abstraction also cannot be too high; otherwise, the learned domain would be useless since it could only generate nonspecific and broad test scenarios. Those would be very difficult to execute automatically.

Precisely, we must map all numeric variables, arrays, strings, and custom data structures to predicates. This step is required because we cannot automatically decide what is necessary to describe all game-play rules fully.
Interestingly, mapping all available variables is both too much detail and not necessarily sufficient at the same time. To completely model the behavior of a general computer program, we would also need to represent the ``hidden'' data like stack traces, the instruction pointer, the variables in the game engine, the operating system, and properties that are implied by the level design or interactions in the physics engine that are not represented in the code. For example, it is possible to jump from one location to another.

Furthermore, we would like to (1-to-1) map planning actions to certain significant functions and methods in the code that represent player actions. The reason is that we want to call these methods from the plans obtained by the learned model.
This step is problematic since STRIPS actions have a very limited structure. On the other hand, functions in code are usually much more complex. They contain branching, loops, other function calls, recursion, and other elements a STRIPS action cannot do. Therefore, mapping most functions from the program code one to one is impossible with the planning actions. 
One solution would be to break up game-code functions into simpler elements that can be mapped to planning actions. We would also need to use additional helper predicates to express the stack traces, program pointers, etc. The disadvantage is that the generated plans would then contain these elementary actions instead of the high-level ones we desired.

Alternatively, we could achieve a ``one to many'' mapping where one function from the program code gets mapped to a couple of planning actions, each representing the whole original function. This step can be done easily and automatically, but from our experience, many very specific planning actions are learned, and the model does not generalize well. It is somewhat similar to the situation known as \emph{overfitting} in machine learning.

We could mitigate these problems by supporting additional PDDL features like conditional effects and universal quantifiers so that more code functions can be mapped to a single (or at least fewer than before) planning action(s). This process is complicated since most action model learning methods only support STRIPS-like action definitions.

\section{Conclusion}

In this paper we presented a workflow that applies research from the areas of automated planning and action model learning in order help with the automation of regression testing during video game development. We demonstrated it with a proof-of-concept project and discussed what challenges need to be addressed to further automate its steps.

\subsection{Future Work}
% In future work 
We plan to evaluate and refine the workflow on larger and more complex games. We are currently experimenting by applying it on a real-time strategy game. We also want to run proper user studies to quantify the usefulness of the proposed workflow compared to just creating the domains manually. Furthermore, we want to address and resolve the challenges related to further automating the synthesis of PDDL domains mentioned in the previous section.

\bibliography{main,gaming}

\end{document}